\documentclass[10pt]{article}
\usepackage{fullpage,citesort,epsfig,graphics,amsbsy}
\usepackage{psfrag}
\usepackage[normal,small]{caption}
\newcommand{\beq}{\begin{equation}}
\newcommand{\eeq}{\end{equation}}
\newcommand{\bea}{\begin{eqnarray}}
\newcommand{\eea}{\end{eqnarray}}

\newcommand{\be}{\begin{equation}}
\newcommand{\ee}{\end{equation}}
\newcommand{\bq}{\begin{eqnarray}}
\newcommand{\eq}{\end{eqnarray}}

\newcommand{\ie}{{\it i.e.\ }}
\def\math{\mathsurround=0pt }
\def\leftrightarrowfill{$\math \mathord\leftarrow \mkern-6mu \cleaders\hbox{$\mkern-2mu \mathord- \mkern-2mu$}\hfill
 \mkern-6mu \mathord\rightarrow$}
\def\overleftrightarrow#1{\vbox{\ialign{##\crcr
     \leftrightarrowfill\crcr\noalign{\kern-1pt\nointerlineskip}
     $\hfil\displaystyle{#1}\hfil$\crcr}}}

\newcommand{\bfs}{\boldsymbol}

\let\l=\lambda

 \def\bd{\begin{document}} \def\ed{\end{document}}
\def\ds{\documentstyle} \let\fr=\frac \let\bl=\bigl \let\br=\bigr
\let\Br=\Bigr \let\Bl=\Bigl
\let\bm=\bibitem
\let\na=\nabla
\let\pa=\partial \let\ov=\overline
\def\ft#1#2{{\textstyle{{\scriptstyle #1}\over {\scriptstyle #2}}}}
\def\fft#1#2{{#1 \over #2}}
\def\vp{\varphi}
\def\sst#1{{\scriptscriptstyle #1}}
\def\oneone{\rlap 1\mkern4mu{\rm l}}
\def\td{\tilde}
\def\wtd{\widetilde}
\def\dalemb#1#2{{\vbox{\hrule height .#2pt
        \hbox{\vrule width.#2pt height#1pt \kern#1pt
                \vrule width.#2pt}
        \hrule height.#2pt}}}
\def\square{\mathord{\dalemb{6.8}{7}\hbox{\hskip1pt}}}
\def\wtd{\widetilde}
\def\R{\rlap{\rm I}\mkern3mu{\rm R}}
\def\im{{\rm i}}
\def\tilg{\tilde{g}}
\def\tilF{\tilde{F}}
\def\tilA{\tilde{A}}
\def\varf{\varphi}
\def\tilf{\tilde{\phi}}
\def\tilh{\tilde{h}}
\def\rme{{\rm e}}
\def\ep{\epsilon}
\def\0{{(0)}}
\def\9{{(9)}}
\def\8{{(8)}}
\def\7{{(7)}}
\def\6{{(6)}}
\def\5{{(5)}}
\def\4{{(4)}}
\def\3{{(3)}}
\def\2{{(2)}}
\def\1{{(1)}}
\newcommand{\trace}{{\rm Tr}}
\newcommand{\ub}{\overline{U}}
\newcommand{\vb}{\overline{V}}
\newcommand{\uh}{\widehat{U}}
\newcommand{\vh}{\widehat{V}}
\newcommand{\ubh}{\overline{\widehat{U}}}
\newcommand{\vbh}{\overline{\widehat{V}}}
\newcommand{\lb}{\bar{\l}}
\newcommand{\Fb}{\overline{F}}
\newcommand{\Fh}{\widehat{F}}
\newcommand{\Fbh}{\overline{\widehat{F}}}
\newcommand{\Ab}{\overline{A}}
\newcommand{\Ah}{\widehat{A}}
\newcommand{\Abh}{\overline{\widehat{A}}}
\newcommand{\Gb}{\overline{G}}
\newcommand{\Gh}{\widehat{G}}
\newcommand{\Gbh}{\overline{\widehat{G}}}
\newcommand{\Pb}{\overline{P}}
\newcommand{\Ph}{\widehat{P}}
\newcommand{\Pbh}{\overline{\widehat{P}}}
\newcommand{\Qb}{\overline{Q}}
\newcommand{\Qh}{\widehat{Q}}
\newcommand{\Qbh}{\overline{\widehat{Q}}}
\newcommand{\Bb}{\overline{B}}
\newcommand{\Bh}{\widehat{B}}
\newcommand{\Bbh}{\overline{\widehat{B}}}
\newcommand{\fhns}{\hat{F}^{\rm (NS)}}
\newcommand{\fhrr}{\hat{F}^{\rm (RR)}}
\newcommand{\ahns}{\hat{A}^{\rm (NS)}}
\newcommand{\ahrr}{\hat{A}^{\rm (RR)}}
\newcommand{\hhrr}{\hat{H}^{\rm (RR)}}
\newcommand{\hchi}{\hat{\chi}}
\newcommand{\hphi}{\hat{\phi}}
\newcommand{\htau}{\hat{\tau}}
\newcommand{\cG}{{\cal G}}
\newcommand{\cGb}{\overline{{\cal G}}}
\newcommand{\cH}{{\cal H}}
\newcommand{\cP}{{\cal P}}
\newcommand{\cPb}{\overline{{\cal P}}}
\newcommand{\cQ}{{\cal Q}}
\newcommand{\cQb}{\overline{{\cal Q}}}
\newcommand{\cM}{{\cal M}}
\newcommand{\cN}{{\cal N}}
\newcommand{\cO}{{\cal O}}
\newcommand{\cD}{{\cal D}}
\newcommand{\cL}{{\cal L}}
\newcommand{\cA}{{\cal A}}
\newcommand{\cB}{{\cal B}}
\newcommand{\hg}{\hat{g}}
\newcommand{\cE}{{\cal E}}

\newcommand{\vpp}{\mbox{$\langle{\scriptstyle++}\rangle$}}
\newcommand{\vmp}{\mbox{$\langle{\scriptstyle-+}\rangle$}}
\newcommand{\vppp}{\mbox{$\langle{\scriptstyle+++}\rangle$}}
\newcommand{\vmpp}{\mbox{$\langle{\scriptstyle-++}\rangle$}}
\newcommand{\vpmp}{\mbox{$\langle{\scriptstyle+-+}\rangle$}}

\begin{document}
\setlength{\captionmargin}{20pt}
\begin{titlepage}
\begin{flushright}
UFIFT-HEP-04-07\\
hep-th/0405018
\end{flushright}

\vskip 3cm

\begin{center}
\begin{Large}
{\bf Renormalization of Quantum Fields on 
the\\ Lightcone Worldsheet I: Scalar Fields\footnote{Supported 
in part by the Department
of Energy under Grant No. DE-FG02-97ER-41029. 
}}
\end{Large}

\vskip 2cm
{\large 
 Charles B. Thorn\footnote{E-mail  address: {\tt thorn@phys.ufl.edu}}
}
\vskip0.20cm
{\it Institute for Fundamental Theory\\
Department of Physics, University of Florida,
Gainesville FL 32611}


\vskip 1.0cm
\end{center}

\begin{abstract}\noindent
We show that the lightcone worldsheet formalism, constructed to
represent the sum of the bare planar diagrams of scalar
$\phi^3$ field theory, survives the renormalization
procedure in space-time dimensions $D\leq6$. Specifically
this means that all the counter-terms, necessary to produce a successful
renormalized perturbation expansion to all orders, can
be represented as local terms in the lightcone worldsheet
action. Because the worldsheet regulator breaks Lorentz invariance, we find the
need for two non-covariant counter-terms, in addition to the
usual mass, coupling and wave function renormalization. One
of these can be simply interpreted as a rescaling of transverse
coordinates with respect to longitudinal coordinates. The second
one introduces couplings between the matter and ghost
worldsheet fields on the boundaries.

\end{abstract}
\vfill
\end{titlepage}
\section{Introduction}
The possible duality between string and field theory
has been a recent active topic of investigation
since Maldacena proposed that IIB superstring theory on
an AdS$_5\times$S$_5$ background is equivalent to
supersymmetric Yang-Mills field theory with
extended ${\cal N}=4$ supersymmetry \cite{maldacena}. 
The most exciting application of this new string
theory technique is, without doubt, the confinement problem of QCD.
Most mechanisms proposed to account for quark confinement
involve the formation of a color flux tube or gluon chain, and
it may well be that formulating QCD as a string theory provide the
most tractable realization of such a physical mechanism.
Most of the literature on this 
subject concentrates on how typically
field theoretic phenomena are accounted for in
a string formulation \cite{klebanovs,polchinskis}. The program
initiated by Bardakci and me takes the opposite tack, seeking a direct 
construction of a stringy worldsheet formalism that
sums the planar diagrams of quantum field theory
\cite{bardakcit,thornsheet,gudmundssontt}. These
three articles lay the foundations of this 
``bottom-up'' approach: the first sets up a worldsheet
formalism for scalar $\Phi^3$ theory, the second
for Yang-Mills theory, and the third for the whole 
range of interesting supersymmetric Yang-Mills theories,
including ${\cal N}=1,2,4$ extended supersymmetry. 
We note that in the AdS/CFT context, an approach
connecting perturbative field theory to string theory on pp-wave
backgrounds \cite{berensteinmn} shares to some extent the
spirit and goals of our program, as does Witten's formulation of perturbative
gauge theory as string on twistor space \cite{wittentwistorstring}.

The worldsheet formalism developed 
in \cite{bardakcit,thornsheet,gudmundssontt} 
reproduces the multi-loop expansion of the {\it bare} planar
Feynman diagrams, singled out by 't Hooft's large
$N_c$ limit \cite{thooftlargen}. Because the bare diagrams
are divergent in interesting space-time dimensions (\ie $D>2$), 
the representation must be regarded as formal except when
applied to tree diagrams. It is therefore crucial to
establish that the worldsheet description survives
the renormalization procedure which systematically
removes the ultraviolet infinities of quantum field theory. 
Since the worldsheet formalism sacrifices manifest Lorentz invariance,
it is logically possible that counter-terms required to
ensure Lorentz invariance after renormalization might
not have a local worldsheet interpretation. Such an
outcome would significantly diminish the usefulness of
the formalism for accessing non-perturbative physics in the
field theory. The purpose of this article is to
tackle this issue for the case of scalar field theory. Our
main result is that the worldsheet picture indeed survives
renormalization: all necessary counter-terms have a
local worldsheet interpretation for the scalar case.
We defer the corresponding study of gauge theories and supersymmetric
field theories to future articles.

The lightcone worldsheet \cite{goddardgrt} is naturally based on light-cone
coordinates: an imaginary time $\tau=ix^+=i(t+z)/\sqrt2$
in the range $0\leq\tau\leq T$
and a worldsheet spatial coordinate $0\leq\sigma\leq p^+$
chosen so that the $p^+$ density is uniform. The worldsheet
path integral representation of 
an arbitrary planar diagram is built from the
worldsheet path integral for
a single scalar propagator \cite{bardakcit}, 
which in the massless case, is simply
\begin{eqnarray}
\exp\left\{-{T{\boldsymbol p}^2\over2p^+}\right\}
&=&\int DcDbD{\boldsymbol q}\ e^{-S}\\
S&=& \int_0^T d\tau \int_0^{p^+}d\sigma\left[{1\over2}{\bfs q}^{\prime2}
-{\bfs b}^\prime{\bfs c}^\prime\right]
\label{wsprop}
\end{eqnarray}
where Dirichlet boundary conditions, ${\boldsymbol q}(0,\tau)
={\boldsymbol q}_1$, ${\boldsymbol q}(p^+,\tau)
={\boldsymbol q}_2$, with ${\boldsymbol q}_2-{\boldsymbol q}_1
={\boldsymbol p}$, are imposed on the bosonic worldsheet variables
${\bfs q}(\sigma,\tau)$. 
The Grassmann variables ${\bfs b}(\sigma,\tau),{\bfs c}(\sigma,\tau)$
are also subject to Dirichlet conditions: ${\bfs b}(0,\tau)=
{\bfs c}(0,\tau)={\bfs b}(p^+,\tau)={\bfs c}(p^+,\tau)=0$.
The worldsheet formalism maps
every planar diagram to a worldsheet with several
internal boundaries each at fixed values of $\sigma$.
The length and location of each of these internal boundaries
are of course integrated over the whole worldsheet.
The worldsheet target space fields ${\boldsymbol q}(\sigma,\tau)$
satisfy Dirichlet conditions ${\boldsymbol q}={\boldsymbol q}_i$
on the $i$th boundary, and each ${\boldsymbol q}_i$
is integrated (the momenta flowing through the propagators
of the multi-loop diagram are obtained from the
${\boldsymbol q}_i$ as the various differences ${\boldsymbol q}_i-
{\boldsymbol q}_j$). The ghosts satisfy ${\bfs b}={\bfs c}=0$
on each internal boundary.

Since the focus of this article is the renormalization
of individual Feynman diagrams, the fine details of the worldsheet 
construction will not play a central role except to dictate
a regularization of the diagrams which enables a
rigorous passage from worldsheet picture to conventional
Feynman diagram and {\it vice versa}. For its rigorous 
definition the worldsheet formalism relies on a discretization
of $\sigma,\tau$, and hence of $T_i,p_i^+$ the light cone 
times and longitudinal momenta associated with the
various propagators of the diagram. On the other hand,
conventional Feynman diagrams require continuous $T_i,p_i^+$.
The ultraviolet divergences of quantum field theory
correspond in lightcone variables to infinities due to
integration at large transverse momentum. These transverse
momentum infinities will
get entangled with, and will spoil, the continuum limit of
the $T_i,p_i^+$ unless they are regulated independently of these 
longitudinal variables \cite{glazek,thorngote}.
The requirement that this transverse regulator be local on the
worldsheet then dictates that it be applied only to the
boundary values ${\bfs q}_i$. Such a cutoff is local
in both $\sigma,\tau$ because it only need be applied
at the beginning or end of an internal boundary (because of
the Dirichlet conditions), i.e.  at a point on the worldsheet. 
It is particularly convenient for
our analysis to simply impose a Gaussian cutoff, i.e. to
insert in the integrand the factor $e^{-\delta\sum{\bfs q}^2_i}$
\cite{bardakcitimp,bardakci,thorngote}.
This factor can be directly interpreted as a local modification of the
worldsheet action.

With $\delta>0$ and fixed, the rigorously defined world sheet path integral
for each multi-loop planar
diagram can be explicitly evaluated on the worldsheet lattice \cite{gilest} 
and then the continuum limit of the $T_i, p_i^+$ for the various
propagators can be safely taken. The result, essentially 
by construction, reduces to
one of the standard representations of the Feynman diagram
in momentum space with the regulator factor $e^{-\delta\sum{\bfs q}^2_i}$
inserted in the integrand. Because the $\delta$
regulator is in place, these integrals are manifestly finite.
In Section 2 we describe how this reduction takes place. 

After this reduction, there remains an almost conventional
analysis of the renormalization procedure in the context of this
somewhat novel regulator. The novelty stems from the
fact that the $q_i$'s, the variables subject to the cut-off, 
are not the momenta 
flowing through the propagators. Rather, they are ``dual-momentum'' 
variables, 
one assigned to each loop. There is also a set of external
dual-momenta $q^e_i$, one assigned to each region between
external lines.
The momentum flowing through the propagator
that separates loop $i$ from loop $j$ is the difference $q_i-q_j$.
Thus the regulator breaks a ``translation'' symmetry
$q^e_i\to q^e_i+a$ enjoyed by the {\it bare} unregulated 
diagram\footnote{Because the regulator only 
cuts off the transverse components
of $q$, the translation symmetry in the longitudinal momenta remains 
unbroken.}. Because of this broken symmetry, with $\delta>0$
the $n$-point function depends on $n$ dual-momenta rather than on $n-1$ 
actual momenta. Formally the limit $\delta\to0$ should restore
the symmetry and the amplitudes should become independent of
one of the dual-momenta. Because of ultraviolet
divergences, the introduction of counter-terms is necessary to ensure 
that this happens.

In Section 3 we describe the properties of this regularization in
great detail. In section 4 we discuss the self-energy and
its renormalization through two loops by direct calculation.
Then an all orders argument is constructed in Section 5.
The one loop three point vertex is calculated in Section 6,
and the correct asymptotically free behavior is confirmed.
Finally in the concluding Section 7 we return to the worldsheet
formalism and show how the new counter-terms required by
renormalization can be represented locally on the worldsheet.

\section{From the Worldsheet to the Schwinger Representation}
\label{sec2}

In this article we will not need much of the detailed
worldsheet formalism, which is rigorously defined on
a worldsheet lattice. But for the reader's convenience we
review the worldsheet construction for scalar field theory in an appendix.  
By construction, the evaluation of the
worldsheet path integral representing a specific
planar Feynman diagram produces a certain discretized version of
the usual multi-loop integral. Each propagator appears
in its mixed ${\bfs p},p^+>0,x^+$ representation
\bea
\int {dp^-\over2\pi}e^{-ix^+p^-}{-i\over p^2+\mu_0^2-i\epsilon}
&=&{\theta(x^+)\over 2p^+}e^{-ix^+({\bfs p}^2+\mu_0^2)/2p^+}\nonumber\\
&\to&{\theta(\tau)\over 2p^+}e^{-\tau({\bfs p}^2+\mu_0^2)/2p^+}.
\eea
The Feynman integration is 
over all independent $\tau_i,p^+_i,{\bfs p}_i$. However the
worldsheet lattice formalism produces instead sums over
discretized $\tau_i=k_ia,p^+_i=l_im$, while keeping the ${\bfs p}_i$
integrals continuous. However, in the presence of the regulator $\delta>0$
described in the introduction, one can safely replace all
of the discretized sums by continuous integrals.

We would like to now show that these perhaps unfamiliar lightcone
multi-loop integrals are identical to the covariant
Feynman integrals in which each propagator is written in a
Schwinger representation. 
\bea
{1\over p^2+\mu^2}=\int_0^\infty dT e^{-T(p^2+\mu^2)}.
\label{schwingerrep}
\eea
Indeed, it is straightforward to show that the number of
independent $\tau_i,p^+_i$ in the diagram's
lightcone representation is
precisely equal to the number of $T_i$ in the diagram's 
Schwinger representation. If one explicitly carries out the
Gaussian integrals in the two representations by completing the
square the remaining integrals in the two representations
will be of the same dimensionality. The integrands are very
similar except that the determinant prefactor from the lightcone
is raised to the $(D-2)/2$ power compared to the $D/2$ power
in the Schwinger representation. One can make
the exponentials in the
integrands identical by changing integration variables
from the $\tau_i,p^+_i$ to appropriate $T_i$. It then turns out
that the Jacobian for this change of variables supplies the
missing determinant factors.

To understand why this happens, just consider the transform
to light-cone representation of the Schwinger representation:
\bea
-i\int {dp^-\over2\pi}e^{-ix^+p^-}\int idT e^{-iT({\bf p}^2
+\mu_0^2-2p^+p^--i\epsilon)}
&=&-i\int idT\delta(x^+-2p^+T) e^{-iT({\bf p}^2+\mu_0^2)}\nonumber\\
&\to&\int dT\delta(\tau-2p^+T) e^{-T({\bf p}^2+\mu_0^2)}.
\eea
From this result we see that the appropriate change of
variables is $T=\tau/2p^+$. It is interesting and satisfying
that the passage to imaginary $x^+$ in the lightcone
representation is completely equivalent to writing the
Schwinger representation with a real exponential, which
of course is only meaningful after the Wick rotation to
Euclidean space.

For the rest of the discussion of renormalization we need 
no longer refer to the explicit worldsheet representation.
We only have to write the usual covariant rules using
dual momenta $q_i$, and insert the regulator factor
$e^{-\delta\sum_i{\bfs q}_i^2}$. Once we have established
the form of the counter-terms required for renormalization
we shall return to give their worldsheet representation
at the end of the article.

\section{Regularization}
Draw an arbitrary planar diagram so that its lines divide the
plane into different regions, the external lines
all going off to infinity. Then the external lines
bound infinite regions, and the finite regions fill each
loop of the multi-loop diagram.  For each loop introduce
a momentum $q^\mu_i$, assigned to the loop's region. Then each
propagator line separates two regions, say $i_1$ and $i_2$,
and the propagator's momentum is then taken to be
$q_{i_1}-q_{i_2}$, and momentum is automatically conserved.
We regulate each diagram by including in the integrand the
factor $e^{-\delta\sum_{i=1}^L{\boldsymbol q}_i^2}$. Since we are
using a light-cone world sheet we only cut off the transverse
momentum integrals, because we want to preserve
longitudinal boost invariance\footnote{One 
could easily extent the cutoff to the
longitudinal variables, but then the light-cone interpretation
would be obscured.}. This regularization sacrifices full Lorentz
invariance, but respects the $O(D-2)$ rotational invariance in transverse
space as well as the longitudinal boost invariance.
The transverse boost invariances generated by $M^{\pm i}$ are broken, 
and it will require counter-terms to restore 
them in the renormalizable case of 6 space-time dimensions.
The task of the worldsheet renormalization program is to enumerate all of
the necessary counter-terms and show that each has a local worldsheet
interpretation. 

Without loss of generality, we can and do restrict attention to
proper (\ie connected one particle irreducible) diagrams,
with propagators removed from external legs. Such diagrams
never have tadpole sub-diagrams, which would be problematic
for the lightcone description (because $p^+>0$), though not for a covariant
description. The only 1PIR diagram involving a tadpole is
the one-point function itself, $\langle\Phi\rangle$. 
It is true that the lightcone description
has no convenient representation of the one point function.
However, in a covariant description, the only effect of 
tadpoles as sub-diagrams in larger (improper) diagrams is pure
mass renormalization, which means their effect can be absorbed
in an additive constant in the self-energy counter-term.
In this article we assume that this is always done and therefore
drop tadpoles completely. Then we can freely
pass back and forth between light-cone and covariant descriptions, 
as long as we refrain from considering
the one-point function itself. Since the one-point function cannot
be directly measured in any case, this is no limitation
on the lightcone description. If needed, the value of the
one-point function can be related via the field equations 
to $\langle\Phi^2\rangle$,
which in turn can be extracted from the high momentum
limit of the two point function.

It is convenient to employ the Schwinger representation
of each propagator (\ref{schwingerrep}):
\bea
{1\over p^2+\mu^2}=\int_0^\infty dT e^{-T(p^2+\mu^2)}\nonumber
\eea
which enables the execution of all loop momentum integrals
by completing the square in the exponents of the Gaussian integrals.
To describe this for an $L$ loop diagram, assemble the
loop momenta in an $L$ dimensional vector $q$ and call $M_0$ the 
$L\times L$ symmetric matrix that describes the quadratic
terms in the $q_i$, so the exponent reads
\bea
-q^T\cdot(M_0+\delta)q+v^T\cdot q +q^T\cdot v -B
\label{diagrambilinear}
\eea
where the $L$-vector $v$ describes the couplings to the momenta
assigned to the external regions and $B$ is a bilinear form in those
external momenta. It is understood that $\delta\neq0$ only for the
transverse components. Then the result of the loop integrations is just
\bea
&&{\pi^L\over\det M_0}\left({\pi^L\over\det(\delta+ M_0)}\right)^{(D-2)/2}
\exp\left\{-B+v^T\cdot {1\over\delta+M_0}v\right\}=\nonumber\\
&&{\pi^L\over\det M_0}\left({\pi^L\over\det(\delta+ M_0)}\right)^{(D-2)/2}
\exp\left\{-B+v^T\cdot {1\over M_0}v
-\delta {\boldsymbol v}^T{1\over M_0}\cdot{1\over\delta+M_0}
{\boldsymbol v}\right\}
\label{diagramint}
\eea
We see that the shift of $M_0$ by $\delta$ regulates the integration
region near the zeroes of the determinant, which is the source
of ultraviolet divergences in the diagram. The first two terms
in the exponent are manifestly Lorentz invariant
and are precisely what they would be in the 
unregulated theory. The last term in the exponent breaks Lorentz invariance
because it depends explicitly on the transverse momentum
components. If we could argue that it were negligible (as $\delta\to0$), 
we could assert from the
known proofs of renormalizability that all divergences 
as $\delta\to0$ could
be covariantly absorbed in the renormalization of mass $\mu$ and coupling $g$
to all orders in perturbation theory.

The term in question 
is nominally of order $O(\delta)$ but since it also
depends on the $T_i$'s we must check this estimate more carefully.
First note that $q_0\equiv (\delta+M_0)^{-1}{\boldsymbol v}$ is in fact
the location of the minimum of a bilinear form in the $q_i$'s
that has the interpretation as the potential energy of 
$L$ particles tied to each other and to the fixed
external momenta with a bunch of springs
with spring constants $T_i>0$ and to the origin with springs of
spring constant $\delta$. It is obvious that the resulting
equilibrium has every $q_{0 i}$ within the simplex with vertices
at the origin and the external momenta. If $\delta=0$ they are
within the simplex with vertices at the external momenta. In 
either case it follows that $|{\boldsymbol q}_{0i}|$ is
uniformly bounded by the largest external momentum. Thus we can conclude
that the term in question is uniformly bounded over the
whole integration region by $L\delta |{\boldsymbol q_{ext}}|_{max}^2$. 
Thus the $O(\delta)$ estimate is rigorous.

Even so, Lorentz non-covariance can survive due to ultraviolet divergences
of degree $1/\delta$ or worse which can overwhelm
the $O(\delta)$ suppression. Fortunately, in a renormalizable theory
we can isolate where these divergences can occur and accordingly
identify the subtractions necessary to remove these contributions
which would violate Lorentz invariance. Indeed the ultraviolet
divergences in vertex parts are only logarithmic in $\delta$ while those in 
self-energy parts are $O(1/\delta)$. Thus we can focus on the
self-energy divergences, but of course we must follow their impact
as sub-diagrams in larger diagrams as well. Our strategy will be first
to directly analyze one and two loop diagrams and then to
develop a recursive argument that the subtraction procedure
works to all orders.

Before turning to the renormalization procedure, let us 
establish rigorous limitations on how non-covariant effects
can contribute to a general regulated bare diagram. We can
expand the non-covariant factor in powers of $\delta$
\bea
\exp\left\{-\delta {\boldsymbol v}^T{1\over M_0}\cdot{1\over\delta+M_0}
{\boldsymbol v}\right\}
=\sum_{n=0}{1\over n!}\left(-\delta {\boldsymbol v}^T{1\over M_0}
\cdot{1\over\delta+M_0}{\boldsymbol v}\right)^n
\eea
For a fixed diagram with $L$ loops, this series will terminate
at a finite $n\leq L$ when $\delta\to0$, because the worst divergence one
will encounter is a $1/\delta$ for each self-energy insertion
and the number of independent self energy insertions is
obviously bounded by $L$, the number of loops. Thus we can conclude
that, at worst, the explicit dependence on the transverse momenta
is a polynomial of order $\leq L$ with (possibly divergent)
but Lorentz invariant coefficients. Furthermore, since 
transverse $O(D-2)$ rotational invariance is maintained by
the regularization the polynomial must be a rotational scalar. 

The divergences in skeleton diagrams are primitive, coming entirely
from the region of integration where all propagator
momenta are large, and given by the superficial divergence of the
diagram as all internal momenta get large together. In the
renormalizable case ($D=6$), 3-point function skeletons are
log divergent and will be covariant as $\delta\to0$. 
The only skeleton self-energy diagram is the one loop bubble,
which has a $1/\delta$ divergence, so non-covariance in it as
$\delta\to0$ resides in a quadratic polynomial in the
$q$'s. If, following Ward, we 
develop a diagrammatic expansion of the derivative of the
self energy, $\partial\Pi(q,q^\prime)/\partial q^\mu$, there
will be an infinite number of skeleton contributions.
Because $\Pi$ is symmetric under $q\leftrightarrow q^\prime$
it is uniquely determined by this derivative up to an additive
constant.
The skeletons
contributing to this derivative of the self energy are linearly
divergent and their non-covariant contributions will be simply linear
in the transverse momenta. Furthermore, since the divergences 
in skeletons require all momenta to be large, the
coefficients of these polynomials must be constants.

Let us summarize the situation for skeleton diagrams.
The derivative of the self energy is linearly divergent:
both its infinity and $\delta$ artifacts are canceled by
a counter-term of the form 
\bea
{\partial\over\partial q^\mu}\Pi_{\rm C.T.}=
{\partial\over\partial q^\mu}
\left(-\alpha{\bfs q}^2-\beta({\bfs q}-{\bfs q}^\prime)^2
+(Z^{-1}-1)(q-q^\prime)^2\right)
\eea
which implies by symmetry
\bea
\Pi_{\rm C.T.}=
\delta\mu^2-\alpha({\bfs q}^2+{\bfs q}^{\prime2})
-\beta({\bfs q}-{\bfs q}^\prime)^2
+(Z^{-1}-1)\left((q-q^\prime)^2+\mu^2\right).
\eea
The three point skeletons are logarithmically divergent and
will possess no $\delta$ artifacts. The infinity is canceled 
by a constant counter-term
\bea
\Gamma^3_{\rm C.T.}=gZ^{-3/2}(Z_1-1).
\eea
All higher point skeletons are finite.

On the other hand the infinities and
$\delta$ artifacts caused by
divergences in sub-diagrams can be higher order
polynomials than the superficial divergence indicates, and
the coefficients can depend non-analytically on the 
Lorentz invariants constructed from the external momenta.
Indeed, powers of transverse momenta higher than the
superficial divergence must, for dimensional
reasons, be multiplied by reciprocal powers of momenta or mass,
implying non-analyticity. In a conventional regularization
scheme, divergent sub-diagrams are automatically taken
care of by the renormalization subtractions performed
on primitive diagrams at lower orders of perturbation theory.
Since the $\delta$ regularization is not quite of the conventional
type, we need to show that this procedure continues to properly handle
sub-diagram artifacts in spite of its novel features.
 
\begin{figure}
\begin{center}
\psfrag{'q'}{}
\psfrag{'qprime'}{}
\psfrag{'q1'}{}
\psfrag{'q2'}{}
\psfrag{'q3'}{}
\includegraphics[width=4cm]{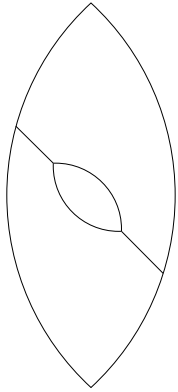}
\caption{A three loop diagram.}
\end{center}
\label{threeloop}
\end{figure}

The necessary counter-terms established by consideration of 
skeleton diagrams can be incorporated in the theory by choosing the 
bare propagator
\bea
\Delta_0(q,q^\prime)&=&{1\over (q-q^\prime)^2+\mu_0^2+\alpha({\boldsymbol q}^2
+{\boldsymbol q}^{\prime2})+\beta({\boldsymbol q}-{\boldsymbol q}^\prime)^2}
\label{modbareprop}
\eea
and by renormalizing the bare parameters $\mu_0,g_0$.
The parameters $\alpha,\beta$ parameterize violations of Lorentz invariance in
the zeroth order theory which are to be tuned to cancel the violations
generated by residual artifacts as $\delta\to0$. To examine how this
works in perturbation theory, we might attempt to assume
$\alpha=\beta=0$ at zero coupling and expand the bare propagator in
powers of $\alpha,\beta$. However such a na\"ive approach
runs into the difficulty that there are regions of momentum integration
where ${\bfs q}^2,{\bfs q}^{\prime2}\gg\mu^2,(q-q^\prime)^2$,
and in these regions the $n$th term in the
expansion of the propagator about $\alpha=\beta=0$ blows up
like ${\bfs q}^{2n}/\mu^{2n}$, ruining the power counting
behind renormalizability. This difficulty first arises in three-loop
diagrams such as Fig.~\ref{threeloop} which is the lowest order
diagram with a self-energy insertion on a line that does not
border the planar diagram.
Note that this issue does not arise in conventional
regularizations because then the propagator, including potential
counter terms, depends only on the momentum differences, so powers
of momentum in the numerator are always compensated by corresponding
powers in the denominator. The same compensation will occur in the
$\delta$ regularization provided that 
we study renormalization with $\alpha\to
\epsilon>0$ at zero coupling. Then the renormalized 
parameters of the theory are $\mu,g,\epsilon$ all held fixed
as $\delta\to0$, while $\mu_0,g_0,\beta,\alpha-\epsilon$ are
adjusted so that infinities and $\delta$ artifacts are canceled
order by order in perturbation theory in $g$. Only at the
very end of the calculation, after renormalization and
after the limit $\delta\to0$ do we set $\epsilon=0$ and recover
full Lorentz invariance\footnote{Organizing the perturbative expansion
in this way corresponds to the way a nonperturbative
evaluation of the worldsheet path integral would be done. Namely, one
would evaluate a physical quantity at finite $\delta$ and fixed
$\alpha, \beta, g_0, \mu_0$. Then one would study how the scaling
behavior as $\delta\to0$ depends on these bare parameters searching
for parameter sets that give Lorentz invariant predictions.
Another way to organize the perturbative calculation, which would
automatically give the usual Lorentz invariant result, would be to
send $\delta\to0$ inside out, starting with the ``innermost'' 
integrals together with their counter-terms before doing the
next outer loop integration. This would
automatically reproduce the standard BPHZ subtraction procedure.
We need to deal with the situation where all momentum integrals 
and subtractions are done at finite $\delta$, and $\delta\to0$
afterwards. Different ways of organizing the $\delta\to0$ limit
must give the same results in terms of physical renormalized 
parameters, but they need not, and in general will not, give the
same relations of physical parameters to bare parameters.}. 
To identify the self energy counter-term we
introduce the renormalized mass $\mu^2=\mu_0^2+\delta\mu^2$ and allow
for wave function renormalization $Z$. 
Then expanding the bare propagator we find 
\bea
\Delta_0(q,q^\prime,\epsilon)&=& 
\sum_{n=0}^\infty{Z^{n+1}\over[(q-q^\prime)^2+\mu^2
+\epsilon({\boldsymbol q}^{2}+{\boldsymbol q}^{\prime2})]^{n+1}}
(\Pi_{\rm C.T.})^n\\
\Pi_{\rm C.T.}&=&{\delta\mu^2}+\left({1\over Z}-1\right)
\left[(q-q^\prime)^2+\mu^2\right]
-\left({\alpha}-{\epsilon\over Z}\right)
({\boldsymbol q}^{2}+{\boldsymbol q}^{\prime2})
-{\beta}({\boldsymbol q}-{\boldsymbol q}^\prime)^2
\label{cteps}
\eea

Keeping $\epsilon\neq0$ modifies the analysis of $\delta$ dependence
slightly. The matrix $M_0$ in (\ref{diagrambilinear}) is changed, 
when acting on the
transverse variables, to 
$M_\epsilon\equiv M_0+\epsilon M^{\rm diag}_0$ where $M^{\rm diag}$
means the matrix obtained by dropping the off-diagonal elements
of the matrix $M$. Also the bilinear in external momenta
$B$ acquires a term linear in $\epsilon$, $B\to B_\epsilon
=B+\epsilon B_\perp$ where $B_\perp$ is obtained from $B$
by dropping the terms in longitudinal momenta. Then the
result of doing the momentum integrations (\ref{diagramint})
now reads:
\bea
&&{\pi^L\over\det M_0}
\left({\pi^L\over\det(\delta+ M_\epsilon)}\right)^{(D-2)/2}
\exp\left\{-B_\epsilon+v^T\cdot {1\over\delta+M_\epsilon}v\right\}
=\nonumber\\
&&{\pi^L\over\det M_0}\left({\pi^L\over\det(\delta+ M_\epsilon)}
\right)^{(D-2)/2}
\exp\left\{-B_\epsilon+v^T\cdot {1\over M_\epsilon}v
-\delta {\boldsymbol v}^T{1\over M_\epsilon}\cdot{1\over\delta+M_\epsilon}
{\boldsymbol v}\right\}
\eea
where it is understood that $M_\epsilon$ is replaced by $M_0$
for the longitudinal components. The argument that the $\delta$
dependent term in the exponent is uniformly of order $O(\delta)$ over the
whole integration region remains valid. Thus as $\delta\to0$
that term only gives contributions when multiplying divergences
of order $1/\delta$ or worse which can only come from self energy insertions.
Of course with $\epsilon\neq0$, removing these contributions 
by suitable counter-terms will not restore Lorentz covariance.
But all of the residual non-covariance will disappear when
we set $\epsilon=0$ at the end of the calculation.

\section{Self Energy for $\Phi^3$}
In order to acquaint the reader with some of the novelties
of calculations using the $\delta$ regulator, we  carry out
in this section a direct calculation or the self energy through two loops,
with an explicit separation of all divergences and 
Lorentz-violating artifacts. 
Since keeping $\epsilon>0$ is not essential through two loop
order, we shall for simplicity set $\epsilon=0$ for 
most of the calculation. However, as an 
example to show that $\epsilon\neq0$ causes no special 
difficulties, we shall show the one loop calculation with $\epsilon>0$.
We call the bare self-energy $\Pi_0$, but it is convenient
to calculate $Z\Pi_0$ and absorb the factor of $Z$ in the bare coupling 
by defining the renormalized coupling $g=g_0Z^{3/2}/Z_1$, where
$Z_1$ is the three vertex renormalization constant. In other words we
write down the Feynman rules in terms of renormalized mass and coupling,
canceling infinities against 
the self-energy counter-term $Z\Pi_{\rm C.T.}$ (see
(\ref{cteps}) and the three
vertex counter-term $g(Z_1-1)\Phi^3$, which are included in
the Feynman rules, rather than absorbing them in redefinitions
of the bare parameters.
  
\subsection{One Loop}
The unsubtracted one-loop self-energy diagram has the value
\bea
Z\Pi_0&=&{g^2\over(4\pi)^3}\int_0^\infty{dT_1dT_2
\over(T_1+T_2)(T_1+T_2+\delta)^2}\nonumber\\
&&\qquad\qquad\exp\left\{-\mu^2(T_1+T_2)-{T_1T_2\over T_1+T_2}(q-q^{\prime})^2
-\delta{(T_1{\boldsymbol q}+T_2{\boldsymbol q}^\prime)^2
\over(T_1+T_2)(T_1+T_2+\delta)}\right\}\nonumber\\
Z\Pi_0&=&
{g^2\over(4\pi)^3}\int_0^\infty{dT e^{-\mu^2 T}\over (T+\delta)^2}\int_0^1 dx
\exp\left\{-Tx(1-x)(q-q^{\prime})^2
-{\delta T\over T+\delta}(x{\boldsymbol q}+(1-x){\boldsymbol q}^\prime)^2
\right\}
\eea
The quadratic divergence can be simply extracted with an integration
by parts, and the log divergence isolated by one further subtraction
in the remaining integrand
\bea
Z\Pi_0&=&-{g^2\over(4\pi)^3}
\int_0^\infty{dT\over T+\delta}\int_0^1 dx\left[
H_0(e^{-HT}-e^{-\mu^2T})+{\delta^2(x{\boldsymbol q}
+(1-x){\boldsymbol q}^\prime)^2
\over (T+\delta)^2}(e^{-HT}-1)\right]\nonumber\\
&&+{g^2\over(4\pi)^3}\left\{{1\over\delta}-
\int_0^1 dx\left[H_0I(\mu^2\delta)
+{1\over2}(x{\boldsymbol q}
+(1-x){\boldsymbol q}^\prime)^2\right]\right\}\\
H&\equiv&\mu^2+x(1-x)(q-q^{\prime})^2
+{\delta\over T+\delta}(x{\boldsymbol q}+(1-x){\boldsymbol q}^\prime)^2
\equiv H_0+{\delta\over T+\delta}(x{\boldsymbol q}
+(1-x){\boldsymbol q}^\prime)^2\\
I(t)&\equiv&\int_0^\infty {e^{-ut}du\over 1+u}\quad{}_{\widetilde{t\to0}} 
\quad\ln{1\over t}
\eea
We can choose the first order contribution of the 
self energy counter term (\ref{cteps}), $Z\Pi^{(1)}_{\rm C.T.}$, to cancel the
second line, which will then cancel the divergences as well
as the Lorentz violating term which would survive the $\delta\to0$ limit. 
In particular we find for the wave function renormalization
to this order,
\bea
Z^{(1)}&=&1-{1\over6}{g^2\over(4\pi)^3}I(\mu^2\delta)
\label{zoneloop}
\eea
Thus we write the one-loop 
renormalized self energy (at finite $\delta$) as
\bea
\Pi^{(1)}&\equiv&Z(\Pi_0+\Pi^{(1)}_{\rm C.T.})\nonumber\\
&=&-{g^2\over(4\pi)^3}
\int_0^\infty{dT\over T+\delta}\int_0^1 dx\left[
H_0(e^{-HT}-e^{-\mu^2T}) +{\delta^2(x{\boldsymbol q}
+(1-x){\boldsymbol q}^\prime)^2\over (T+\delta)^2}(e^{-HT}-1)
\right]
\eea
As long as $\delta>0$ Lorentz invariance is clearly violated by the
explicit dependence of this expression on the transverse components of the
momenta. However, because of the subtracted counter-term these violations
disappear in the limit $\delta\to0$ at {\it fixed} $q,q^\prime$.
Indeed, in this limit the second term in square brackets 
$O(\delta)$ and the expression $HT$ in the first term
tends to $H_0T+O(\delta\ln\delta)$, so we have
\bea
\Pi_1&\sim&-{g^2\over(4\pi)^3}
\int_0^\infty{dT\over T}\int_0^1 dx H_0(e^{-H_0T}-e^{-\mu^2T})
={g^2\over(4\pi)^3}\int_0^1 dx H_0 \ln{H_0\over\mu^2}
\eea
which is obviously Lorentz invariant and finite.

Before moving on to two loops, we pause to
indicate how $\epsilon\neq0$
complicates the one loop calculation.
Then we use the more general
free propagator
\bea
{Z\over \mu^2+(q-q^\prime)^2+\epsilon({\boldsymbol q}^2
+{\boldsymbol q}^{\prime2})},
\eea
associated with the self energy counter-term (\ref{cteps}).
Now the
bare unsubtracted one-loop self energy is
\bea
Z\Pi_0&=&{g^2\over(4\pi)^3}\int_0^\infty{dT_1dT_2
\over(T_1+T_2)[(T_1+T_2)(1+\epsilon)
+\delta]^2}\exp\left\{-\delta{(T_1{\boldsymbol q}
+T_2{\boldsymbol q}^\prime)^2
\over(1+\epsilon)(T_1+T_2)((T_1+T_2)(1+\epsilon)+\delta)}\right\}\nonumber\\
&&\qquad\qquad\exp\left\{-\mu^2(T_1+T_2)-{T_1T_2\over T_1+T_2}(q-q^{\prime})^2
-\epsilon[T_1{\boldsymbol q}^2+T_2{\boldsymbol q}^{\prime2}]
-{\epsilon\over1+\epsilon}{(T_1{\boldsymbol q}
+T_2{\boldsymbol q}^\prime)^2
\over T_1+T_2}\right\}\nonumber\\
Z\Pi_0&=&
{g^2\over(4\pi)^3}\int_0^\infty{dT \over (T(1+\epsilon)+\delta)^2}
\int_0^1 dx\exp\left\{-TH_\epsilon
-{\delta T\over (1+\epsilon)(T(1+\epsilon)+\delta)}(x{\boldsymbol q}
+(1-x){\boldsymbol q}^\prime)^2
\right\}\\
H_\epsilon&=&\mu^2+x(1-x)(q-q^{\prime})^2+\epsilon[x{\boldsymbol q}^2
+(1-x){\boldsymbol q}^{\prime2}]
+{\epsilon\over1+\epsilon}{(x{\boldsymbol q}
+(1-x){\boldsymbol q}^\prime)^2}
\eea
As before, integration by parts and a subtraction in the remaining integrand
isolates the divergences:
\bea
Z\Pi_0&=&\nonumber\\
&&\hskip-1cm-{g^2\over(4\pi)^3}
\int_0^\infty{dT\over (1+\epsilon)(T(1+\epsilon)+\delta)}
\int_0^1 dx\left[H_\epsilon(e^{-HT}-e^{-\mu^2T})
+{\delta^2(x{\boldsymbol q}
+(1-x){\boldsymbol q}^\prime)^2\over 
(1+\epsilon)(T(1+\epsilon)+\delta)^2}(e^{-HT}-1)\right]\nonumber\\
&&+{g^2\over(4\pi)^3}\left\{{1\over(1+\epsilon)\delta}-
\int_0^1 dx\left[H_\epsilon{I({\mu^2\delta/(1+\epsilon)})
\over(1+\epsilon)^2} 
+{1\over2(1+\epsilon)^3}(x{\boldsymbol q}
+(1-x){\boldsymbol q}^\prime)^2\right]\right\}\\
H&\equiv&H_\epsilon
+{\delta(x{\boldsymbol q}+(1-x){\boldsymbol q}^\prime)^2
\over (1+\epsilon)(T(1+\epsilon)+\delta)}
\eea
We choose $\Pi^{(1)}_{\rm C.T.}$ to cancel the third line obtaining
for the renormalized self energy for $\delta,\epsilon>0$
\bea
\Pi_1&=&\nonumber\\
&&\hskip-1.5cm-{g^2\over(4\pi)^3}
\int_0^\infty{dT\over (1+\epsilon)(T(1+\epsilon)+\delta)}
\int_0^1 dx\left[H_\epsilon(e^{-HT}-e^{-\mu^2T})
+{\delta^2(x{\boldsymbol q}
+(1-x){\boldsymbol q}^\prime)^2\over 
(1+\epsilon)(T(1+\epsilon)+\delta)^2}(e^{-HT}-1)\right]
\eea
and (\ref{zoneloop}) changes to
\bea
Z_\epsilon^{(1)}&=&1-{1\over6}{g^2\over(4\pi)^3}{1\over(1+\epsilon)^2}
I\left({\mu^2\delta\over1+\epsilon}\right)
\eea
Taking the limit $\delta\to0$ yields
\bea
\Pi_1&\sim&-{g^2\over(4\pi)^3}
\int_0^\infty{dT\over(1+\epsilon)^2 T}
\int_0^1 dx H_\epsilon(e^{-H_\epsilon T}-e^{-\mu^2T})
={1\over(1+\epsilon)^2}{g^2\over(4\pi)^3}\int_0^1 dx H_\epsilon 
\ln{H_\epsilon\over\mu^2}
\eea
a finite result, but of course it is not Lorentz invariant 
unless $\epsilon=0$. Note that there is no
particular sensitivity to $\epsilon$ in this limit.
The presence of $\epsilon\neq0$ does
not change the qualitative power-counting: the self energy grows
quadratically with momenta and the propagator falls
off quadratically, any number of self energy
insertions in a skeleton diagram does not alter the
high momentum behavior of the integrand mod logs. 
The advantage of keeping $\epsilon>0$
is that this power counting also holds for $\delta\neq0$.

\subsection{Two Loops}
\begin{figure}
\begin{center}
\psfrag{'a'}{(a)}
\psfrag{'b'}{(b)}
\psfrag{'c'}{(c)}
\includegraphics[width=5in]{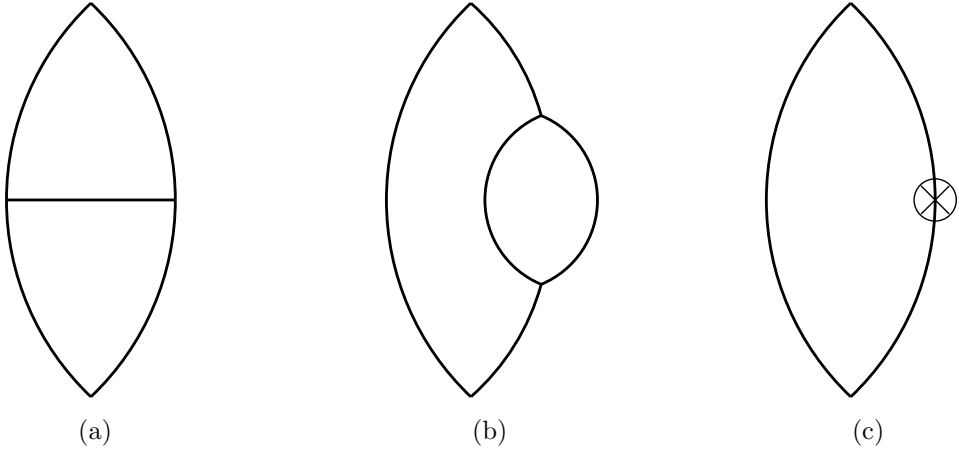}
\end{center}
\caption{Self energy two loop diagrams}
\label{twoloopgraphs}
\end{figure}
Now we turn to the two loop self-energy, setting $\epsilon=0$
for simplicity.
First consider the two loop diagram in Fig.\ref{twoloopgraphs}a).
The matrix $M_0$ and related quantities for this graph are given by
\bea
M_0&=&\pmatrix{T_1+T_3+T_5&-T_5\cr -T_5& T_2+T_4+T_5\cr}\\
{1\over\delta+M_0}&=&{1\over\det(M_0+\delta)}
\pmatrix{\delta+T_2+T_4+T_5&T_5\cr T_5& \delta+T_1+T_3+T_5\cr}\\
\det(M_0+\delta)&=&(\delta+T_1+T_3+T_5)(\delta+T_2+T_4+T_5)-T_5^2
\eea
The vector describing the external momenta is given by
\bea
v=\pmatrix{T_1q+T_3q^\prime\cr T_2q+T_4q^\prime\cr}
\eea
so we find
\bea
{1\over\delta+M_0}v={1\over\det(M_0+\delta)}
\pmatrix{q(T_1\delta+T_1(T_2+T_4+T_5)+T_2T_5)
+q^\prime(T_3\delta+T_3(T_2+T_4+T_5)+T_4T_5)\cr
q(T_2\delta+T_2(T_1+T_3+T_5)+T_1T_5)
+q^\prime(T_4\delta+T_4(T_1+T_3+T_5)+T_3T_5)}
\eea
This two loop amplitude in Schwinger representation is
\bea
&&\hskip-1cm\Pi_{2,a}^0={g^4\over(4\pi)^6}\int_0^\infty \prod_{i=1}^5 dT_i  
{1\over\det M_0}{1\over\det(M_0+\delta)^2}
\exp\left\{-\mu^2\sum_{i=1}^5 T_i-(q-q^\prime)^2{N(T_i)\over\det M_0}
-\delta{\boldsymbol v}^T\cdot{1\over M_0}{1\over\delta+M_0}
{\boldsymbol v}\right\}\\
&&\hskip-1cm N(T_i)=T_1T_2(T_3+T_4)+T_3T_4(T_1+T_2)+T_5(T_1+T_2)(T_3+T_4)
\eea
To analyze the $\delta\to0$ limit we can expand to first order in the
last term of the exponent since it is $O(\delta)$ and the worst
divergence in the integration is $O(1/\delta)$. Moreover,
for this diagram that divergence comes only when all the $T_i$
are $O(\delta)$. Thus the second term of this expansion can
be simplified by setting the exponent to zero and scaling all the
$T_i\to\delta T_i$, exploiting the homogeneity in the
$T_i$ of the integrand:
\bea
{\rm Second~Term}\to-{g^4\over(4\pi)^6}\int_0^\infty \prod_{i=1}^5 dT_i  
{1\over\det M_0}{1\over\det(M_0+1)^2}
\ {\boldsymbol v}^T\cdot{1\over M_0}{1\over1+M_0}{\boldsymbol v}
\eea
which is clearly a quadratic polynomial in the transverse
components of the external momenta with constant and finite coefficients,
which is of the form of the allowed counter-terms for the self-energy.
The first term in the expansion is a manifestly covariant
regularization of the standard two loop amplitude. It contains the
usual logarithmic overlapping divergences that renormalize the
vertices of the one loop self energy as well as the overall
quadratic $1/\delta$ divergence that is to be absorbed in the mass,
as well as and overall log divergence that contributes to the
two loop wave function renormalization. The overall quadratic divergence
is easy to extract by dropping the exponent and scaling the $T_i$:
\bea
{\rm Quad~Divergence}={1\over\delta}{g^4\over(4\pi)^6}
\int_0^\infty \prod_{i=1}^5 dT_i{1\over\det M_0}{1\over\det(M_0+1)^2}.
\eea
The overlapping log divergences correspond either to $T_1,T_3,T_5\sim0$
$T_2,T_4$ finite or $T_2,T_4,T_5\sim0$ with $T_1,T_3$ finite. 
For definiteness take the first case. We note the simplifications
\bea
N(T_i)&\to& (T_1+T_3+T_5)T_2T_4\nonumber\\
\det(M_0+\delta)&\to&(\delta+T_1+T_3+T_5)(T_2+T_4+\delta)\nonumber
\eea
So the integration over small $T_1,T_3,T_5$ involves
\bea
\int_{T_1+T_3+T_5<\Delta} dT_1dT_3dT_5
{1\over (T_1+T_3+T_5)(\delta+T_1+T_3+T_5)^2}
={1\over2}\int_0^{\Delta}dT{T\over (T+\delta)^2}
\sim{1\over2}\ln{\Delta\over\delta}
\eea
and the corresponding contribution to the self energy is
\bea
{\rm Overlap135}={g^4\over(4\pi)^6}{1\over2}\ln{\Delta\over\delta}
\int_\epsilon^\infty dT_2dT_4{e^{-\mu^2(T_2+T_4)}
\over (T_2+T_4)(T_2+T_4+\delta)^2}
\exp\left\{-(q-q^\prime)^2{T_2T_4\over T_2+T_4}
\right\}.
\eea
This is precisely the vertex renormalization needed to renormalize the charge
at one vertex of the one loop self energy. The 245 overlap
takes care of the renormalization at the other vertex.

We next take up the two loop diagram shown in Fig.\ref{twoloopgraphs}b).
We list the corresponding
ingredients of the answer: 
\bea
M_0&=&\pmatrix{T_1+T_2+T_3+T_4&-T_4\cr -T_4& T_4+T_5\cr}\\
{1\over\delta+M_0}&=&{1\over\det(M_0+\delta)}
\pmatrix{\delta+T_4+T_5&T_4\cr T_4& \delta+T_1+T_2+T_3+T_4\cr}\\
\det(M_0+\delta)&=&(\delta+T_1+T_2+T_3+T_4)(\delta+T_4+T_5)-T_4^2
\eea
The vector describing the external momenta is
\bea
v=\pmatrix{T_1q+(T_2+T_3)q^\prime\cr T_5q^\prime\cr}
\eea
so we find
\bea
{1\over\delta+M_0}v={1\over\det(M_0+\delta)}
\pmatrix{qT_1(\delta+T_4+T_5)
+q^\prime[(T_2+T_3)(\delta+T_4+T_5)+T_4T_5]\cr
qT_1T_4
+q^\prime[T_5(\delta+T_1+T_2+T_3+T_4)+T_4(T_2+T_3)]}
\eea
This two loop amplitude in Schwinger representation is
\bea
&&\hskip-1cm\Pi_{2,b}^0={g^4\over(4\pi)^6}\int_0^\infty \prod_{i=1}^5 dT_i  
{1\over\det M_0}{1\over\det(M_0+\delta)^2}
\exp\left\{-\mu^2\sum_{i=1}^5 T_i-(q-q^\prime)^2{N(T_i)\over\det M_0}
-\delta{\boldsymbol v}^T\cdot{1\over M_0}{1\over\delta+M_0}
{\boldsymbol v}\right\}\\
&&\hskip-1cm N(T_i)=T_1T_4T_5+T_1(T_2+T_3)(T_4+T_5)
\eea
Again it is sufficient expand to first order in the
last term of the exponent since it is $O(\delta)$ and the worst
divergence in the integration is $O(1/\delta)$. However,
for this diagram there is, in addition to an overall
$1/\delta$ divergence coming from the region when all the $T_i$
are $O(\delta)$, a $1/\delta$ divergence from the
region $T_4,T_5=O(\delta)$. As before we can scale
all the $T_i$ by $\delta$, obtaining:
\bea
{\rm Second~Term}&\to&\nonumber\\
&&\hskip-1in-{g^4\over(4\pi)^6}\int_0^\infty \prod_{i=1}^5 dT_i  
{1\over\det M_0}{1\over\det(M_0+1)^2}
\ {\boldsymbol v}^T\cdot{1\over M_0}{1\over1+M_0}{\boldsymbol v}
\exp\left\{-\mu^2\delta\sum_{i=1}^5 T_i
-\delta(q-q^\prime)^2{N(T_i)\over\det M_0}\right\}
\label{seinsertionnoncov}
\eea
In these variables, the exponent must be retained to converge the
integral in the region $T_1,T_2,T_3\gg T_4,T_5$. In this region, without the
exponent the
$T_4,T_5$ integration is linearly convergent but the $T_1,T_2,T_3$
integration is log divergent at infinity. 
Because $T_4,T_5$ integration converges in any case we can simplify
the exponent by neglecting $T_4,T_5$ compared to $T_1,T_2,T_3$. 
\bea
{\rm Exponent}\to-\mu^2\delta(T_1+T_2+T_3)-\delta(q-q^\prime)^2
{T_1(T_2+T_3)\over T_1+T_2+T_3}
\eea
The prefactors
to the exponential in this limit have the behavior
\bea
{1\over\det M_0}{1\over\det(M_0+1)^2}
{\boldsymbol v}^T\cdot{1\over M_0}{1\over1+M_0}{\boldsymbol v}
&\sim& {1\over T_{13}T_{45}(T_{13}+1)^2(T_{45}+1)^2}\nonumber\\
&&\left[
{(T_1{\boldsymbol q}
+T_{23}{\boldsymbol q}^\prime)^2\over T_{13}(T_{13}+1)}
+{T_{45}\over T_{45}+1}(x{\boldsymbol q}_1
+(1-x){\boldsymbol q}^\prime)^2\right]
\eea
The shorthand notation used
here is $T_{rs}\equiv\sum_{i=r}^s T_i$. Also $x=T_4/T_{45}$,
and ${\boldsymbol q}_1$ is the $\delta\to1$ limit of
\bea
{\boldsymbol q}_\delta&=&{T_1{\boldsymbol q}+
T_{23}{\boldsymbol q}^\prime\over T_{13}+\delta}\nonumber
\eea
In this region the integration over $T=T_{45}$ at fixed $x$ 
can be easily performed. We change variables $dT_4dT_5=dx TdT$,
and find
\bea
\int dT_4dT_5{1\over\det M_0}{1\over\det(M_0+1)^2}
{\boldsymbol v}^T\cdot{1\over M_0}{1\over1+M_0}{\boldsymbol v}
&\sim& \nonumber\\
&&\hskip-1.5in{1\over T_{13}(T_{13}+1)^2}\left[
{(T_1{\boldsymbol q}
+T_{23}{\boldsymbol q}^\prime)^2\over T_{13}(T_{13}+1)}
+{1\over2}\int_0^1dx(x{\boldsymbol q}_1
+(1-x){\boldsymbol q}^\prime)^2\right]
\eea

The log divergence at large $T_1,T_2,T_3$ also will bring
dependence on the invariant $(q-q^\prime)^2$ into the
coefficients of the Lorentz violating polynomial in the
transverse momenta. Thus the diagram of Fig.\ref{twoloopgraphs}b)
by itself cannot be made covariant by our allowed subtractions.
However, the self energy
insertion is accompanied with a counter-term, corresponding to
the diagram in Fig.\ref{twoloopgraphs}c). We shall show
that the sum of the two diagrams
Fig.\ref{twoloopgraphs}b+c) {\it can} be made Lorentz invariant.

The cross in the counter-term
diagram represents the counter-term insertion required
for Lorentz invariance of the one loop self energy:
\bea
{\rm Insertion}=-{g^2\over(4\pi)^3}\left[{1\over\delta}
-{1\over2}\int_0^1 dx
(x{\boldsymbol q}_1+(1-x){\boldsymbol q}^\prime)^2\right]
\eea
Here $q_1$ is the loop momentum. Evaluating the counter-term diagram
gives
\bea
\Pi_{2\rm C.T.}&=&-{g^4\over(4\pi)^6}\int_0^\infty {dT_1 dT_2 dT_3\over
T_{13}(T_{13}+\delta)^2}\left[{1\over\delta} -{1\over4(T_{13}+\delta)}
-{1\over2}\int_0^1dx(x{\boldsymbol q}_\delta
+(1-x){\boldsymbol q}^\prime)^2\right]
\nonumber\\
&&\exp\left\{-\mu^2T_{13}-(q-q^\prime)^2
{T_1(T_{23})\over T_{13}}-{\delta(T_1{\boldsymbol q}+
(T_{23}){\boldsymbol q}^\prime)^2\over T_{13}(T_{13}+\delta)}
\right\}
\eea
Again we can expand to first order in the last term of the exponent
and the second term in the expansion only survives multiplying
the first two terms in square brackets. Collecting all the
terms quadratic in transverse momenta and scaling out $\delta$:
\bea
\Pi^{\rm Quad}_{2~{\rm C.T.}}&=&
-{g^4\over(4\pi)^6}\int_0^\infty {dT_1 dT_2 dT_3\over
T_{13}(T_{13}+1)^2}\exp\left\{-\mu^2\delta T_{13}-(q-q^\prime)^2\delta
{T_1T_{23}\over T_{13}}\right\}\nonumber\\
&&\left[-\left({1} -{1\over4(T_{13}+1)}\right)
{(T_1{\boldsymbol q}+
T_{23}{\boldsymbol q}^\prime)^2\over T_{13}(T_{13}+1)}
-{1\over2}\int_0^1dx(x{\boldsymbol q}_1+(1-x){\boldsymbol q}^\prime)^2\right]
\eea
Just as before the exponent must be retained to converge the
integration at large $T_1,T_2,T_3$. However the term coming from the
$-1/4/(T_{13}+1)$ in the parentheses within the square brackets
is convergent without the exponent. By inspection, we can see that
all the other terms will exactly cancel the large $T_1,T_2,T_3$ behavior
of the integrand of (\ref{seinsertionnoncov}), providing the
extra convergence needed to make the exponent negligible.
In summary the non-covariant contribution to the sum of the
diagrams of Fig.\ref{twoloopgraphs}bc) has been shown to be
\bea
\Pi^{\rm noncov}_{2bc}&\to&
-{g^4\over(4\pi)^6}\int_0^\infty {dT_1 dT_2 dT_3\over
4T_{13}^2(T_{13}+1)^4}(T_1{\boldsymbol q}+T_{23}{\boldsymbol q}^\prime)^2
\nonumber\\
&&-{g^4\over(4\pi)^6}\int_0^\infty \prod_{i=1}^5 dT_i  
\left[{1\over\det M_0}{1\over\det(M_0+1)^2}
{\boldsymbol v}^T\cdot{1\over M_0}{1\over1+M_0}{\boldsymbol v}
\right.\nonumber\\
&&\left.-{1\over T_{13}T_{45}(T_{13}+1)^2(T_{45}+1)^2}
\left[
{(T_1{\boldsymbol q}
+T_{23}{\boldsymbol q}^\prime)^2\over T_{13}(T_{13}+1)}
+{T_{45}\over T_{45}+1}(x{\boldsymbol q}_1
+(1-x){\boldsymbol q}^\prime)^2\right]\right]
\label{senoncovbc}
\eea
which, though complicated, is easily seen to be a quadratic
polynomial in the transverse momenta with constant coefficients.

\section{An all orders argument}
\subsection{Preliminaries}
In preparation for an all orders
argument let us reconsider the two loop self-energy diagrams
with an eye to understanding, without detailed
calculation, why subtraction
of a second order polynomial in momenta suffices to remove the regularization
artifacts. For any multi-loop diagram
containing no divergent sub-diagrams, the reason
is transparent. Then the regularization dependence could
only come from the region of loop momentum integration
in which {\it all} internal momenta are large. For this region
one can expand in powers of the external momenta, and by
power counting all powers of external momenta higher
than quadratic will be negligible. Unfortunately the only
self energy diagram with no divergent sub-diagrams is the
one loop one. But then we can apply Ward's trick, and calculate the
derivative of the self energy instead.

The core of the problem then is to understand how
to take care of divergent sub-diagrams. At two loops the only
dangerous diagram is the one
which has $\Pi_1$ inserted on one of the propagator lines.
We can write the bare regulated two-loop diagram plus the
one loop diagram with the counter-term corresponding
to the self-energy insertion as
\bea
\Pi_2^0=g^2\int {d^6q_1\over(2\pi)^6}e^{-\delta{\boldsymbol q}_1^2}
{1\over (q_1-q)^2+\mu^2}
{1\over ((q_1-q^\prime)^2+\mu^2)^2}\Pi_1(q_1,q^\prime,\delta).
\eea
We must now argue that when $\delta\to0$
the Lorentz violations in $\Pi_2^0$
reside exclusively in a quadratic polynomial in the 
${\boldsymbol q},{\boldsymbol q}^\prime$ with constant coefficients.
We first note that the subtractions inherent in the definition
of the insertion $\Pi_1$ guarantee that the Lorentz violations due
to $\Pi_1$ will disappear for any finite region of
the $q_1$ integration. They are only 
present for $q_1>O(1/\delta)$. But for large $q_1$, we can expand
the integrand, including $\Pi_1$
in the external momenta, and integrating over the region $q_1>1/\delta$
will yield a polynomial in all components of the external momenta
with coefficients possibly singular in $\delta$
plus a finite remainder that is Lorentz
invariant. The order of this polynomial is limited by power
counting to the degree of divergence of the self energy namely 2.
Since the regulator preserves both transverse rotational invariance
and longitudinal Lorentz invariance, the polynomial must be
a linear combination of $(q-q^\prime)^2$, ${\boldsymbol q}^2$,
${\boldsymbol q}^{\prime2}$, and $({\boldsymbol q}-{\boldsymbol q}^\prime)^2$.
But this is what was to be proved. Thus one can subtract 
a suitable counter-term from $\Pi_2^0$ to obtain $\Pi_2$ as
a covariant function of $(q-q^\prime)^2$ only. This form
of the argument can be used to prove by induction that
subtraction of a counter-term of the form 
$\alpha(g)({\boldsymbol q}^2+{\boldsymbol q}^{\prime2})
+\beta(g)({\boldsymbol q}-{\boldsymbol q}^\prime)^2$ from $\Pi^0$
will be sufficient to render the self energy, as
well as all proper diagrams Lorentz invariant to all orders in
perturbation theory.

As we carry on beyond two loops, inserting renormalized self energy 
and vertex diagrams as sub-diagrams in larger diagrams is supposed to
take care of the divergences due to sub-integrations. To evaluate
the effect of such insertions on the divergence structure of the
larger diagram, however, we need to know the large momentum behavior
of the renormalized insertions. The usual rule of thumb
is that the renormalized propagators and vertex functions
have the same high momentum behavior as their free counterparts
{\it modulo} logarithms. In a standard covariant
calculational scheme, the self-energy and propagator each depend
on the single invariant $(q-q^\prime)^2$, so this rule of thumb translates
to self energies behaving asymptotically like $(q-q^\prime)^2$ and
propagators like $(q-q^\prime)^{-2}$. This has the consequence that
the insertion of a self-energy on a line does not alter the
asymptotic behavior (mod logs), because each such insertion
comes with one extra propagator so there is a cancellation.
As discussed in the previous section, in order for this
same power counting to be maintained with $\delta>0$, it
is important to hold $\epsilon>0$ throughout the renormalization
procedure. Then positive powers of momentum in the
asymptotics of the self energy will be compensated
by corresponding negative powers from extra propagators,
regardless of which components of the momenta are large.
The zeroth order propagator is then
$[(q-q^\prime)^2+\mu^2+\epsilon({\bfs q}^2 + 
{\bfs q}^{\prime2})]^{-1}$, and the extra terms give the inverse free
propagator the same generic asymptotic behavior as
the self energy. 

The original sum of bare diagrams possessed a symmetry under
the shift of all the $q$ momenta by the same constant:
$q_i\to q_i+a$. The $\delta$ regulator destroys this symmetry,
and the mismatch in asymptotic
behavior between the self energy and the free propagator
when $\epsilon=0$ and $\delta>0$ can be traced to the  
fact that the free propagator possesses this symmetry while
the interactions do not. With $\epsilon>0$
the free propagator no longer possesses this symmetry and the power
counting rules are restored. Our procedure then
is to start with $\delta$ and $\epsilon$
non-zero, for which the free propagator violates both the shift symmetry
and Lorentz symmetry. We then calculate to some finite
order in perturbation theory. Then, after
renormalization and adjustment of  counter-terms, we take $\delta\to0$ 
to obtain a finite result containing no $\delta$ artifacts.
The final step is to set $\epsilon\to0$ to restore shift and Lorentz invariance
and establish agreement with standard renormalized perturbation theory.

\subsection{The argument}
It is convenient, for purposes of formulating the all orders argument,
to dispense with the Schwinger representation and to use the expressions
for the multi-loop diagrams as integrals over the $q$'s only.
We use induction on $n$, the number of loops. Specifically,
the induction hypothesis is that through $n$ loops,
the renormalization of $\mu^2, g$ and adjustment of
$\alpha,\beta$ has
rendered all planar 1PIR amplitudes finite when all
external momenta are much less than
$1/\delta$ in such a way that the dependence on these 
external momenta is finite and Lorentz invariant when $\epsilon=0$.

Now consider the $n+1$ loop planar diagrams, including all of the
counter-terms necessary through $n$ loop order. Our task is to argue that
all new infinities and artifacts that appear in this order
can be absorbed in the new counter-terms allowed in this order.
Consider first the region of $q$ integration in which any of the
propagators bordering the planar diagram carries momentum
much less than the cutoff $1/\sqrt\delta$. For this region of integration
the induction hypothesis is operative, guaranteeing that
no infinities or artifacts are present in this region.

Thus we can turn our attention to the region of $q$ integration where
{\it all} of the propagators bordering the planar diagram
carry momenta on the order $O(1/\sqrt\delta)$. In this region it is
safe to expand these propagators in powers of the external momenta they
depend on. Because the diagram is planar, the external momenta 
appear nowhere else in the integrand. The coefficient
of the $k$th power of external momenta
in this expansion is in order of magnitude a
factor $(\sqrt{\delta})^k$ times the value of the
diagram at 0 external momenta. Since there are a finite number of loops
in the diagram of interest, there is a limit
on the number of inverse powers of $\delta$ in this
coefficient, so this series must terminate and we
conclude that this dangerous region of integration produces at
worst a polynomial in all of the external momenta.
Simple power counting indicates that this integration region
of the self-energy
diagrams supply at most a single power of $1/\delta$, of the
three point functions at most a $\ln\delta$, and of the
four and higher point functions only positive powers of $\delta$.
This conclusion does not hold for the unrenormalized diagrams
because of the possibility of divergent sub-diagrams.
However, including the counter-terms and renormalization through
$n$ loops (which is part of the induction hypothesis) renders these
sub-integrations finite and restores the simple power counting.
It is crucial however that $\epsilon$ is held fixed at
a positive value during this renormalization process.

The upshot is that at $n+1$ loop order the new infinities and
artifacts reside in a quadratic polynomial for the
self energy, in a constant for the three point function
and are absent in the four and higher point functions.
But these are exactly of the form of the allowed new counter-terms
and the argument is complete. 

\section{The triangle graph and one-loop coupling renormalization}
As a final exercise we complete renormalization at one loop
by evaluating the 1PIR one loop correction
to the cubic vertex in the presence of $\epsilon>0$. 
We obtain for the one-loop triangle diagram at 
fixed $\delta, \epsilon$ 
\bea
Z^{3/2}\Gamma^{(1)}_3&=&
{g^3\over (4\pi)^3}
\int_0^\infty {dT_1dT_2dT_3\over 
T_{13}(T_{13}(1+\epsilon)+\delta)^2}
\exp\left\{-{\delta({\boldsymbol q}^{\prime\prime}T_3+{\boldsymbol q}T_1
+{\boldsymbol q}^\prime T_2)^2\over
(1+\epsilon)T_{13}(T_{13}(1+\epsilon)+\delta)}\right\}\nonumber\\
&&\exp\left\{-{\mu^2}T_{13}
-{T_1T_3(q-q^{\prime\prime})^2
+T_1T_2(q-q^\prime)^2
+T_2T_3(q^\prime-q^{\prime\prime})^2\over T_1+T_2+T_3}
\right\}\nonumber\\
&&\exp\left\{-\epsilon({\boldsymbol q}^2T_1
+{\boldsymbol q}^{\prime2} T_2+{\boldsymbol q}^{\prime\prime2}T_3)
-{\epsilon({\boldsymbol q}T_1
+{\boldsymbol q}^\prime T_2+{\boldsymbol q}^{\prime\prime}T_3)^2\over
(1+\epsilon)T_{13}}\right\}
\eea
We see explicitly that the $\delta\to0$ limit produces
a log divergence which can be removed by a single subtraction.
The counter-term that will accomplish this can be taken at first order to be
\bea
Z^{3/2}\Gamma^{(1)}_{\rm C.T.}&\equiv& g(Z_1-1)^{(1)}
=-{g^3\over (4\pi)^3}\int_0^\infty {dT_1dT_2dT_3\over 
T_{13}(T_{13}(1+\epsilon)+\delta)^2}e^{-{\mu^2}T_{13}}\nonumber\\
&=&-{g^3\over 2(1+\epsilon)^2(4\pi)^3}
\left[I\left({\mu^2\delta\over1+\epsilon}\right)
\left(1+{\mu^2\delta\over1+\epsilon}\right)-1\right]\nonumber\\
&\sim&-{g^3\over 2(1+\epsilon)^2(4\pi)^3}\ln{1+\epsilon\over\mu^2\delta}\\
Z_1&=&1-{g^2\over 2(1+\epsilon)^2(4\pi)^3}\ln{1+\epsilon\over\mu^2\delta}
\eea

We can now write the relation between renormalized and bare
coupling
\bea
{g}&=&g_0{Z^{3/2}\over Z_1}\nonumber\\
&=&{g_0}\left(1+{{g_0}^2\over2(1+\epsilon)^2(4\pi)^3}
\ln{1+\epsilon\over\mu^2\delta}
-{3\over2}{{g_0}^2\over6(1+\epsilon)^2(4\pi)^3}
\ln{1+\epsilon\over\mu^2\delta}\right)\nonumber\\
&=&{g_0}\left(1+{{g_0}^2\over256(1+\epsilon)^2\pi^3}
\ln{1+\epsilon\over\mu^2\delta}\right)
\eea
and the Callan-Symanzik beta function is
\bea
\beta(g)\equiv\mu{dg\over d\mu}=-{g^3\over128(1+\epsilon)^2\pi^3}+O(g^5). 
\label{largenbeta}
\eea
For $\epsilon=0$ this is the known result for the beta function
for {\it planar} $\Phi^3$ field theory.

\section{Conclusion: The Renormalized Worldsheet}
In this article we have studied the renormalization
procedure within the regularized setting provided by
the lightcone worldsheet description of the
sum of planar diagrams. We found that, in addition to
counter-terms equivalent to mass, coupling, and
wave function renormalization, two additional
counter-terms must be introduced, parametrized by
$\alpha$ and $\beta$ in the modified bare
propagator (\ref{modbareprop}). We now show how these
additional counter-terms enter the worldsheet
description.

Let us first describe how our regulator $\delta$ modifies
the worldsheet action. The path integrand acquires a factor
$e^{-\delta{\bfs q}_i^2}$ for each internal boundary labeled
by $i$. Because of Dirichlet conditions, we can associate
each such factor with the point on the worldsheet where
the boundary is created. Referring to the worldsheet
lattice, such a point $(i,j)$ is characterized by Ising spin variables
$s_i^j=+1$ and $s_i^{j-1}=-1$ or $P_i^j=1$ and $P_i^{j-1}=0$.
Thus the regulator is incorporated by adding to the worldsheet action the
term
\bea
S_{\rm reg}=\delta\sum_{i,j}{\bfs q}_i^{j2}P_i^j(1-P_i^{j-1}).
\eea

Passing to the mixed $x^+,p^+,{\bfs p}$ representation,
\bea
\Delta_0\to {1\over 2p^+}\exp\left\{-T{(1+\beta)({\bfs q}-{\bfs q}^\prime)^2
+\mu_0^2+\alpha({\bfs q}^2+{\bfs q}^{\prime2})\over2p^+}\right\}.
\label{mixedprop}
\eea
we see that the $\beta$ counter-term  can be included in the worldsheet 
formalism as a simple scale factor multiplying the worldsheet
action
\begin{eqnarray}
\int_0^T d\tau \int_0^{p^+}d\sigma\left[{1\over2}{\bfs q}^{\prime2}
-{\bfs b}^\prime{\bfs c}^\prime\right]&\to&
(1+\beta)\int_0^T d\tau \int_0^{p^+}d\sigma\left[{1\over2}{\bfs q}^{\prime2}
-{\bfs b}^\prime{\bfs c}^\prime\right].
\label{wsaction}
\end{eqnarray}
Physically it describes a renormalization of the scale
of transverse coordinates compared to longitudinal
coordinates, and can also be thought of as a difference
of the speed of light in the transverse directions compared to
the longitudinal direction. Since the regularization we
use breaks the Lorentz symmetry between longitudinal and
transverse directions, it is natural to expect artifacts
in the interacting theory to affect the relative speed of
light in these directions. So in order to end up with Lorentz
symmetry we need to introduce an asymmetry in the bare propagator
tuned to exactly cancel the asymmetry induced by ultraviolet
divergences. From the point of view of the worldsheet 
this adjustment just corresponds to renormalization of the
worldsheet fields, and it changes the structure of the worldsheet action 
very little.

The $\alpha$ counter-term represents a more substantial modification of
the worldsheet action. It is clearly partly a boundary term because
it only involves ${\bfs q},{\bfs q}^\prime$ 
the values of ${\bfs q}(\sigma,\tau)$ on the boundary of the 
propagator's worldsheet. However, it is multiplied in the
exponent by $1/p^+$ which is a global property of the
worldsheet. But this nonlocal feature is shared by the bare
mass term and can be represented in a similar way, by modifying the 
ghost contribution to the action. Recall from \cite{thorntfisheet}
the identity
\begin{eqnarray}
\int \prod_{i=1}^{M-1} {dc_idb_i\over2\pi} \exp\left\{(1+\beta){a\over m}
\left[{b_1c_1\over\eta}+{b_{M-1}c_{M-1}\over\xi}
+\sum_{i=1}^{M-2}(b_{i+1}-b_i)(c_{i+1}-c_i)\right]
\right\}
&=&\nonumber\\
&&\hskip-2in{M\over\eta\xi}\left(1+{\eta+\xi-2\over M}\right)
\left((1+\beta){a\over2\pi m}\right)^{M-1}.
\label{ghostbitsab}
\end{eqnarray}
For $\eta=\xi=1$ the left side of this equation is just the
lattice definition for one time-slice and for one pair
of ghost fields of the worldsheet path integral for the
massless free scalar propagator. Therefore, including
$(D-2)/2=d/2$ ghost pairs and $N$ time-slices and calling the
ghost action, for general $\eta,\xi$, $S^g_{\eta\xi}$,
we can write
\bea
\left(\eta\xi\right)^{Nd/2}\int D{\bfs c}D{\bfs b}\ e^{-S^g_{\eta\xi}}&=&
\left(1+{\eta+\xi-2\over M}\right)^{Nd/2}\int D{\bfs c}D{\bfs b}\ e^{-S^g}
\nonumber\\
&\sim& \exp\left\{-{mTd\over2ap^+}(2-\eta-\xi)\right\}
\int D{\bfs c}D{\bfs b}\ e^{-S^g}\ .
\eea 
Comparing to (\ref{mixedprop}), we see that the bare mass term and the
$\alpha$ counter-term will be produced if we choose $\eta,\xi$
so that
\bea
{md\over a}(2-\eta-\xi)=\mu_0^2+\alpha({\bfs q}^2+{\bfs q}^{\prime2}).
\eea
For example, we could choose
\bea
\eta({\bfs q}^2)=1-{a\over md}(\mu_0^2+\alpha{\bfs q}^2)\ ; 
\qquad\xi({\bfs q}^{\prime2})=
1-{a\over md}\alpha{\bfs q}^{\prime2}
\eea
We see that the $\alpha$ counter-term introduces a non-trivial coupling
between the worldsheet ``matter'' fields and worldsheet ghost fields on the
boundary. To be completely explicit, the term thats must be added
to the Lattice worldsheet action to account for the bare mass and
$\alpha$ counter-terms is
\bea
-S_{\mu\alpha}&=&(1+\beta){a\over m}\sum_{ij}b_i^jc_i^j(1-P_i^j)
\left\{\left({\eta({\bfs q}_{i-1}^{j2})^{-1}}-1\right)
P_{i-1}^j+\left({\xi({\bfs q}_{i+1}^{j2})^{-1}}-1\right)
P_{i+1}^j\right\}\nonumber\\
&&\qquad +\sum_{ij}P_i^j\left\{(1-P_{i+1}^j)\ln\eta({\bfs q}_{i}^{j2})
+(1-P_{i-1}^j)\ln\xi({\bfs q}_{i}^{j2})\right\}.
\eea
The ``bare'' worldsheet action didn't show such a coupling.
It is important to appreciate that in spite of this non-trivial
modification the renormalized worldsheet system retains a
local worldsheet dynamics.

In this article we have developed further the worldsheet model
proposed in \cite{bardakcit} to describe the sum of the planar diagrams of
scalar $\Phi^3$ theory. We have studied the ultraviolet divergence structure
and have established the necessary refinements in the definition
of the model to ensure that it reproduces to all orders the
{\it renormalized} perturbation expansion for spacetime dimensions
$D\leq6$. We saw the need for new counter-terms parametrized by
two new parameters $\alpha, \beta$. At tree level these parameters are 
zero but they must be tuned as a function of coupling and
mass so that the renormalized perturbation theory is correctly
reproduced. Unfortunately, we do not know {\it a priori}
what values to choose for these new parameters.

Our refined worldsheet system is completely regulated
and rigorously defined on a worldsheet lattice. On the
lattice it enjoys a local worldsheet dynamics. 
One is therefore in a position
to apply numerical methods for its solution. In such
studies we must regard $\mu_0,g_0,\alpha,\beta$ as
adjustable parameters of a system that is not
Lorentz invariant for generic values of these
parameters. Thus one must be able to scan over different
parameter sets and test for consistency with Lorentz
invariance to help determine them.  

One promising numerical
approach is Monte Carlo simulation. The worldsheet
path integral for the $q$ variables and the Ising
spins clearly involves a positive definite integrand.
The Grassmann $bc$ variables, introduced to give
a local description, clearly introduce minus signs
and ensuing complications for Monte Carlo methods.
On the other hand, if one integrates out the ghosts,
positivity is again restored at the expense of some
non-locality. Because the ghost dynamics is independent on each time
slice, the relevant determinants are only one dimensional,
and there is hope that their evaluation won't 
be prohibitively time consuming. 

Our real interest in worldsheet methods is in tackling the
large $N_c$ (planar) limit of QCD, not planar $\Phi^3$
scalar field theory. It is therefore an important next step to 
extend the analysis of this article to that case. 
The ``bare'' worldsheet system has already peen constructed
in \cite{thornsheet}. We anticipate that the 
necessary counter-terms will be more numerous than for
the scalar theory studied here, because of the gluon spin and
issues with $p^+=0$ divergences. We hope to resolve these
difficulties in the near future.

\vskip14pt
\noindent\underline{ Acknowledgments}: 
I am grateful to K. Bardakci, S. Glazek, J. Klauder,
and S. Shabanov for valuable discussions. I would also like to thank
M. Shifman and M. Voloshin for their helpful criticism of this work. 
This research was supported in part by the Department
of Energy under Grant No. DE-FG02-97ER-41029.

\appendix
\section{Appendix}
\begin{figure}
\begin{center}
\includegraphics[width=6cm,height=12cm,angle=90]{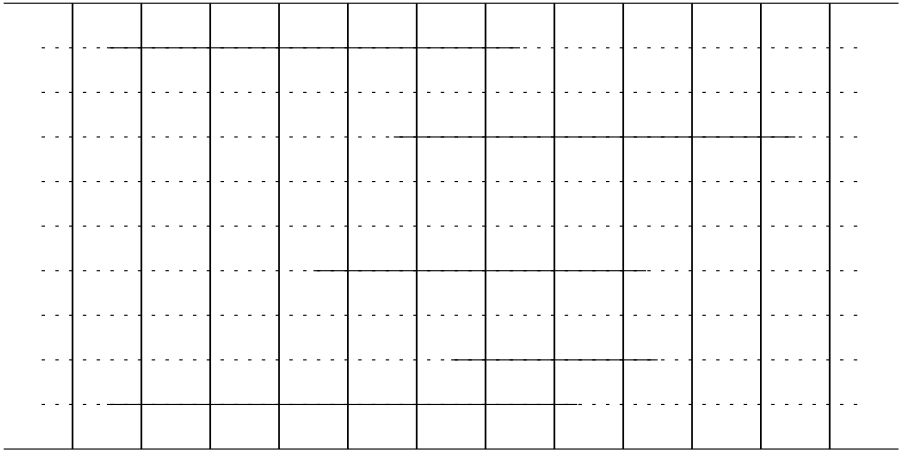}
\end{center}
\caption{Lattice worldsheet for a multi-loop planar diagram. Time
($\tau=ix^+=k a$) flows left to right, and the vertical direction measures 
$p^+=Mm$.
The horizontal solid line segments show the location of the loops.}
\label{worldsheet}
\end{figure}
The worldsheet for the general planar diagram, an
example of which is depicted in Fig.~\ref{worldsheet}, has
an arbitrary number of horizontal solid lines marking the
location of the internal boundaries corresponding
to loops. Each interior link $j,j-1$ of
a solid line at spatial location $k$ requires a factor of
$\delta({\boldsymbol q}_{k}^j-{\boldsymbol q}_{k}^{j-1})$.
To supply such factors, assign an Ising spin $s_k^j
=\pm1$ to each site of the lattice. We assign $+1$ if the site $(k,j)$ 
is crossed by a vertical solid line, $-1$ otherwise. 
We also use the spin up projector $P_k^j=(1+s_k^j)/2$.
We implement the Dirichlet
conditions on boundaries using the Gaussian representation of
the delta function:
\begin{eqnarray}
\left({2\pi m\over a}\right)^{d/2}\delta({\boldsymbol q}_{i}^j
-{\boldsymbol q}_{i}^{j-1})
&=&\lim_{\epsilon\to0}{1\over\epsilon^{d/2}}
\exp\left\{-{a\over2m\epsilon}({\boldsymbol q}_{i}^j-
{\boldsymbol q}_{i}^{j-1})^2\right\},
\label{deltarep2}
\end{eqnarray}
We keep $\epsilon$ finite until the end of the calculation.
Using this device, our formula for
the sum of {\it bare} planar diagrams is \cite{bardakcit,thorntfisheet}:
\begin{eqnarray}
T_{fi}&=&\lim_{\epsilon\to0}
\sum_{s_i^j=\pm1}\int DcDbD{\boldsymbol q}
 \exp\left\{\ln{\hat g}\sum_{ij}{1-s_i^js_i^{j-1}\over2}
-{d\over2}\ln\left(1+\rho\right)\sum_{i,j}P_i^j\right\}
\nonumber\\
&&\exp\left\{-{a\over2m}\sum_{i,j}
{({\boldsymbol q}_{i+1}^j-{\boldsymbol q}_{i}^{j})^2}
-{a\over2m\epsilon}\sum_{i,j}P_i^jP_i^{j-1}
{({\boldsymbol q}_{i}^j-{\boldsymbol q}_{i}^{j-1})^2}
\right\}\label{isingsumeps2}\\
&&\exp\left\{{a\over m}
\sum_{i,j}\left[A_{ij}{\boldsymbol b}^j_{i}{\boldsymbol c}^j_{i}
-B_{ij}{b}_{i}^{j}{c}_{i}^{j}
+C_{ij}({\boldsymbol b}_{i+1}^j-{\boldsymbol b}_i^j)
({\boldsymbol c}_{i+1}^j-{\boldsymbol c}_i^j)
-D_{ij}({b}_{i+1}^j-{b}_i^j)({c}_{i+1}^j-{c}_i^j)\right]\right\}\nonumber\\
A_{ij}&=&{1\over\epsilon}{P}_i^j{P}_i^{j-1}
+{P}_i^{j+1}{P}_i^j-{P}_i^{j-1}{P}_i^j{P}_i^{j+1}
+(1-P_i^j)(P_{i+1}^j+P_{i-1}^j)+\rho(1-P_i^j)P_{i-1}^{j-1}P_{i-1}^j\\
B_{ij}&=&(1-P_i^j)\left(P_{i+1}^jP_{i+1}^{j+1}
(1-P_{i+1}^{j-1})+
P_{i-1}^j{P_{i-1}^{j+1}(1-P_{i-1}^{j-1})}
+P_i^{j-1}P_i^{j-2}P_{i+1}^j\right)\\
C_{ij}&=&(1-P_i^j)(1-P_{i+1}^j)
\\
D_{ij}&=&(1-P_i^j)(1-P_{i+1}^j)P_i^{j-1}P_i^{j-2}
\end{eqnarray}
The parameter $\rho=\mu_0^2a/(dm-\mu_0^2a)$
provides a bare mass $\mu_0$ for the gluon. The 
dimensionless coupling constant
${\hat g}$ appearing in this formula is related to the
conventional dimensionful bare coupling $g_0$ by
\bea
{\hat g}^2={g_0^2\over64\pi^3}\left[{m\over2\pi a}\right]^{(d-4)/2}
\eea
Where $d=D-2$ is the dimensionality of transverse space.
The first exponent in (\ref{isingsumeps2}) supplies a factor of
${\hat g}$ whenever a boundary is created or destroyed.
The second exponent includes the action $S_q$ for the free
propagator together with the exponent in the Gaussian representation
of the delta function that enforces Dirichlet boundary conditions
on the solid lines.
The contents of this exponent
go to the discretized action for the
light-cone quantized string, if the quantity
$a^2P_i^jP_i^{j-1}/m^2\epsilon$ is replaced by
$1/T_0^2$, with $T_0$ the string rest tension.
The third exponent incorporates the $\epsilon$ dependent
prefactor for that representation of the delta
function as a term in the ghost Lagrangian. 
The remaining exponents contain $S_g$ together with
strategically placed spin projectors that arrange the
proper boundary conditions on the Grassmann variables
and supply appropriate $1/p^+$ factors needed at the
beginning or end of solid lines.


\begin{thebibliography}{1}
\bibitem{maldacena}
J. M. Maldacena, {\sl Adv. Theor. Math. Phys.} {\bf 2} (1998) 231-252,
  hep-th/9711200.
\bibitem{klebanovs}
I.~R.~Klebanov and M.~J.~Strassler,
JHEP {\bf 0008} (2000) 052
[arXiv:hep-th/0007191].
\bibitem{polchinskis}
J.~Polchinski and M.~J.~Strassler,
arXiv:hep-th/0003136;
Phys.\ Rev.\ Lett.\  {\bf 88} (2002) 031601
[arXiv:hep-th/0109174];
JHEP {\bf 0305} (2003) 012
[arXiv:hep-th/0209211].
\bibitem{bardakcit}
K.~Bardakci and C.~B.~Thorn, Nucl. Phys. {\bf B626} (2002) 287, hep-th/0110301.
\bibitem{thornsheet}
C.~B.~Thorn,
Nucl.\ Phys.\ B {\bf 637} (2002) 272
[arXiv:hep-th/0203167].
\bibitem{gudmundssontt}
S.~Gudmundsson, C.~B.~Thorn, and T.~A.~Tran,
Nucl.\ Phys.\ B {\bf 649} (2002) 3, 
[arXiv:hep-th/0209102].
\bibitem{berensteinmn}
D.~Berenstein, J.~M.~Maldacena and H.~Nastase,
JHEP {\bf 0204} (2002) 013
[arXiv:hep-th/0202021].
\bibitem{wittentwistorstring}
E.~Witten,
``Perturbative gauge theory as a string theory in twistor space,''
arXiv:hep-th/0312171.
\bibitem{thooftlargen}
G. 't Hooft, {\sl Nucl. Phys.} {\bf B72} (1974) 461.
\bibitem{goddardgrt}
P. Goddard, J. Goldstone, C. Rebbi, and C. B. Thorn, {\sl Nucl. Phys.} {\bf
  B56} (1973) 109.
\bibitem{glazek}
S.~D.~Glazek,
Phys.\ Rev.\ D {\bf 60} (1999) 105030 [arXiv:hep-th/9904029]; 
Phys.\ Rev.\ D {\bf 63} (2001) 116006 [arXiv:hep-th/0012012]; 
Phys.\ Rev.\ D {\bf 66} (2002) 016001 [arXiv:hep-th/0204171]; 
arXiv:hep-th/0307064.
\bibitem{thorngote}
C.~B.~Thorn,
``Fields in the language of string: Divergences and renormalization,''
arXiv:hep-th/0311026.
\bibitem{bardakcitimp}
K.~Bardakci and C.~B.~Thorn,
Nucl. Phys. B {\bf 661} (2003) 235, [arXiv:hep-th/0212254].
\bibitem{bardakci}
K.~Bardakci,
``Self consistent field method for planar phi**3 theory,''
arXiv:hep-th/0308197.
\bibitem{gilest}
R. Giles and C. B. Thorn, {\sl Phys. Rev.} {\bf D16} (1977) 366.
\bibitem{thorntfisheet} 
C.~B.~Thorn and T.~A.~Tran,
Nucl.\ Phys.\ B {\bf 677} (2004) 289, arXiv:hep-th/0307203.
\end{thebibliography}
\end{document}